\documentstyle[12pt,epsf,epsfig,axodraw]{article}
\newcommand{\ie}{{\it i.e.}}
\newcommand{\eg}{{\it e.g. }}
\newcommand{\mrm}[1]{\mbox{\rm #1}}
\newcommand{\beq}{\begin{equation}}
\newcommand{\eeq}{\end{equation}}
\newcommand{\bea}{\begin{eqnarray}}
\newcommand{\eea}{\end{eqnarray}}
\newcommand{\rfn}[1]{(\ref{#1})}

\newcommand{\nn}{\nonumber}

\def\lsim{\mathrel{\vcenter{\hbox{$<$}\nointerlineskip\hbox{$\sim$}}}}
\def\gsim{\mathrel{\vcenter{\hbox{$>$}\nointerlineskip\hbox{$\sim$}}}}

\def\m12{m_{1\!/2}}

\begin{document}
\begin{titlepage}
\pagestyle{empty}
\baselineskip=21pt
\rightline{CERN--TH/2002-133}     
\vskip 0.5in
\begin{center}
{\large{\bf 
Leptogenesis and the Violation of Lepton Number and CP \\ at Low Energies
}}
\end{center}
\begin{center}
\vskip 0.25in
{
{\bf John Ellis}$^{1}$ and 
{\bf Martti Raidal}$^{1,2}$
\vskip 0.15in
{\it
$^1${Theory Division, CERN, CH-1211 Geneva 23, Switzerland}\\
$^2${National Institute of Chemical Physics and Biophysics, 
Tallinn 10143, Estonia}\\ 
}}
\vskip 0.45in
{\bf Abstract}
\end{center}
\baselineskip=18pt \noindent

In the context of the minimal supersymmetric seesaw model, we study the
implications of the current neutrino data for thermal leptogenesis,
$\beta\beta_{0\nu}$ decay, and leptonic flavour- and CP-violating
low-energy observables. We express the heavy singlet-neutrino Dirac Yukawa
couplings $(Y_\nu)_{ij}$ and Majorana masses $M_{N_i}$ in terms of the
light-neutrino observables and an auxiliary hermitian matrix $H$, which
enables us to scan systematically over the allowed parameter space. If the
lightest heavy neutrino $N_1$ decays induce the baryon asymmetry, there
are correlations between the $M_{N_1}$, the lightest active neutrino mass
and the primordial lepton asymmetry $\epsilon_1$ on the one hand, and the
$\beta\beta_{0\nu}$ decay parameter $m_{ee}$ on the other hand. {\it
However, leptogenesis is insensitive to the neutrino oscillation phase}.
We find lower bounds $M_{N_1}\gsim 10^{10}$ GeV for the normal
light-neutrino mass hierarchy, and $M_{N_1}\gsim 10^{11}$ GeV for the
inverted mass hierarchy, respectively, indicating a potentially serious
conflict with the gravitino problem. Depending on $M_{N_1}$, we find upper
(upper and lower bounds) on the lightest active neutrino mass for the
normal (inverted) mass hierarchy, and a lower bound on $m_{ee}$ even for
the normal mass ordering. The low-energy lepton-flavour- and CP-violating
observables induced by renormalization are almost independent of
leptogenesis. The electron electric dipole moment may be close to the
present bound, reaching $d_e\sim 10^{-(27-28)}$ e cm in our numerical
examples, while $d_\mu$ may reach $d_\mu\sim 10^{-25}$ e cm.

\vfill
\vskip 0.15in
\leftline{CERN--TH/2002-133}
\leftline{June 2002}
\end{titlepage}
\baselineskip=18pt


\section{Introduction}

The only convincing experimental evidence for physics beyond the Standard
Model, so far, is provided by neutrino oscillations~\cite{skatm,sno1},
which are generally interpreted as evidence for non-zero neutrino masses.
The smallness of neutrino masses is explained naturally via the seesaw
mechanism~\cite{seesaw} with heavy singlet Majorana neutrinos.  
Leptogenesis scenarios envisage that CP violation in their
out-of-equilibrium decays may induce a non-zero lepton asymmetry, which
may be converted into the observed baryon asymmetry of the Universe via
electroweak sphaleron processes~\cite{fy}.  If, in addition, there is
low-energy supersymmetry as motivated by the hierarchy problem, which is
exemplified by the vastly different mass scales involved in the seesaw
mechanism, low-energy processes violating charged-lepton flavour and CP
may be observable, induced by the neutrino Dirac Yukawa interactions via
renormalization below the heavy-singlet mass scale~\cite{bm,h1}.  In this
scheme, the only source of many observables - neutrino oscillations,
$\beta\beta_{0\nu}$ decay, the baryon asymmetry of the Universe, and
flavour-violating decays of charged leptons and their electric dipole
moments (EDMs) - are the Dirac Yukawa couplings $(Y_\nu)_{ij}$ and the
Majorana masses $M_{N_i}$ of the three heavy singlet neutrinos.

Today, most of our knowledge of $Y_\nu$ and $M_{N_i}$ comes from the
light-neutrino mass and mixing parameters measured in oscillation
experiments, with an additional constraint from searches for $\beta
\beta_{0\nu}$ decays~\cite{bb}.  The oscillation data are converging
towards unique solutions for each of the atmospheric and solar neutrino
anomalies~\cite{sno2}, which represents a major breakthrough in neutrino
physics. On the other hand, the determination of the baryon asymmetry of
the Universe has significantly improved in recent years~\cite{olive}, and
will improve still further with further astrophysical and cosmological
observations, such as those of the MAP and Planck satellites.  In
addition, one expects significant improvements in the future experiments
searching for $\beta\beta_{0\nu}$ decay~\cite{bbfuture},
lepton-flavour-violating (LFV) processes~\cite{Barkov,MECO,nufact} and
electric dipole moments (EDMs)~\cite{BNL,nufact,Lam}. These prospects
motivate us to perform a comprehensive study of these observables in the
minimal seesaw model.

Any study of leptogenesis, neutrino masses, LFV processes and the EDMs of
charged leptons faces the generic difficulty of relating the experimental
information on light neutrino masses and mixings with other observables.
If one takes a top-down approach and fixes $Y_\nu$ and $M_{N_i}$ by some
theoretical or phenomenological argument~\cite{gn}, such as GUT relations,
$U(1)$ or non-Abelian flavour models, phenomenological textures,
democratic principles, arguments of minimal fine-tuning, etc., one can
study the pattern of typical predictions of the model considered, but
cannot perform a comprehensive phenomenological study of the interesting
observables. Even correct numerical consistency with light-neutrino data
is a difficult task in this approach, since it may involve fine-tunings
and must be checked {\it a posteriori}.

These problems can be solved in a bottom-up approach~\cite{di} to neutrino
observables. Parametrizing $(Y_\nu)_{ij}$ and $M_{N_i}$ in terms of the
light neutrino mass matrix ${\cal M}_\nu$ and an auxiliary Hermitian
matrix $H$, as in~\cite{ehrs}, compatibility with the light-neutrino data
is automatic, because ${\cal M}_\nu$ is an input.  In addition, since the
Hermitian matrix $H$ has a physical interpretation as $H=Y_\nu^\dagger D
Y_\nu$, where $D$ is a real diagonal matrix, one has also control over the
renormalization-induced LFV decays. For every $(Y_\nu)_{ij}$ and $M_{N_i}$
generated in this way, one can therefore calculate exactly the weighted
light neutrino mass $m_{ee}$ measured in $\beta\beta_{0\nu}$ 
decay~\cite{bbth}, rates for LFV decays~\cite{h1,nlfv,ci}, and EDMs of charged
leptons~\cite{susyedm,ehrs2}. Moreover, one can also calculate
consistently the leptogenesis CP asymmetries
$\epsilon_i$~\cite{vissani,p}, and, assuming the standard thermal leptogenesis
scenario, also the washout parameters $\kappa_i$~\cite{k-sm,k-mssm,bbp}.  
Hence one can study the correlations between all these parameters and
their dependences on the light and heavy neutrino masses.  To our
knowledge, there has so far been no study of leptogenesis in which {\bf
all} the $\epsilon_i,$ $\kappa_i,$ $m_{\nu_i}$ and $M_{N_i}$ are treated
simultaneously as dynamical variables determined in consistency with the
oscillation data.

In this paper we perform such a comprehensive phenomenological study of
three classes of leptonic observables in the minimal supersymmetric seesaw
model: the effective light-neutrino parameters in ${\cal M}_\nu$, the
baryon asymmetry of the Universe generated via thermal leptogenesis, and
the renormalization induced LFV processes and EDMs of charged leptons.
Assuming the large-mixing-angle (LMA) solution to the solar neutrino
anomaly, we parametrize $Y_\nu$ and $M_{N_i}$ in terms of ${\cal M}_\nu$
and $H.$ In our previous paper \cite{ehrs} we considered only the case
$H=Y_\nu^\dagger D Y_\nu$ where $D_{ii}=\ln (M_{GUT}/M_{N_i}).$ Here we
also consider a different form for $D$ \cite{di}, which yields
 maximal EDMs for the
charged leptons. We assume the standard thermal leptogenesis scenario in
which the observed baryon asymmetry of the Universe is generated only by
the decays of the lightest singlet neutrino $N_1$\footnote{This assumption
is not valid in the case of highly-degenerate heavy neutrinos, but still
holds well in the case of moderate degeneracy~\cite{p}. As the $M_{N_i}$
are output parameters in our approach, it is not suitable for systematic
studies of very degenerate heavy neutrino phenomena.}.  The wash-out
parameter $\kappa_1$ is calculated by solving numerically the analytical
Boltzmann equations of~\cite{bcst}, and correcting by a constant factor in
order to be consistent with the exact results in~\cite{bbp}.
 
We scan randomly over the input parameters: the lightest neutrino mass
$m_{\nu_1}$ (or $m_{\nu_3}$ for the inverted hierarchy of light-neutrino
masses), the two Majorana phases in ${\cal M}_\nu,$ and the entries of the
parameter matrix $H.$ We require the Yukawa couplings $Y_\nu$ to remain
perturbative until the GUT scale, we require the generated baryon
asymmetry to be consistent with observation, and we also require
consistency with all the bounds on LFV processes.

We find interesting correlations between the neutrino observables and
leptogenesis parameters, whilst the LFV processes and EDMs are almost
independent of the leptogenesis constraints.  {\it The influence of the
neutrino oscillation CP phase $\delta$ on leptogenesis is negligible}: the
existence of a baryon asymmetry does not require it to be non-vanishing.
The experimental bound $Y_B\gsim 3\times 10^{-11}$ on the
baryon-to-entropy density ratio $Y_B$ implies the lower bounds
$M_{N_1}\gsim 10^{10}$ GeV and $M_{N_1}\gsim 10^{11}$ GeV for the normally
and inversely ordered light neutrino masses, respectively~\footnote{These
conclusions on neutrino parameters and leptogenesis are valid also in
non-supersymmetric models.}. These bounds put the findings
of~\cite{di2} on rigorous numerical ground, and indicate a serious
potential conflict with the gravitino problem in supergravity 
models~\cite{gravitino}.  
Leptogenesis also implies non-trivial bounds on the mass of lightest light
neutrino.  For the normal mass ordering, there is an $M_{N_1}$-dependent
upper bound on $m_{\nu_1}$, whilst for the inverted hierarchy there are
both upper and lower bounds on $m_{\nu_3}.$ Successful leptogenesis with
$m_{\nu_1}\gsim 0.1$ eV could be allowed for $M_{N_1}\gsim 10^{12}$ GeV.  
There is also an $M_{N_1}$-dependent lower bound on $m_{ee}$ for
normally-ordered light neutrinos, implying its possible measurability in
future experiments. On the other hand, $m_{ee}$ has a preferred value
determined by $\Delta m^2_{atm}$ even in the case of the inverted mass
hierarchy. It tends to be below ${\cal O}(10^{-1})$ eV, making improbable
the discovery of $\beta\beta_{0\nu}$ decay in current experiments.

The rates of LFV processes and EDMs depend also on the soft
supersymmetry-breaking parameters, which we fix by choosing one of the
post-LEP benchmark points~\cite{bench}. We find that $Br(\tau\to\mu
(e)\gamma)$ and $Br(\mu\to e\gamma)$ may easily saturate their present
lower bounds, and that the EDMs of electron and muon may reach $d_e\sim
10^{-(27-28)}$ e cm and $d_\mu\sim 10^{-25}$ e cm, respectively, in our
random samples. We stress in particular that the electron EDM $d_e$ may be
less than an order of magnitude from the present experimental
bound $d_e\lsim 1.6\times 10^{-27}$ e~cm~\cite{eEDM} 
and offers a sensitive probe of the supersymmetric
seesaw model. We find also some correlation between LFV $\tau$ decay rates
and the lower bound on $M_{N_1}$.

The paper is organized as follows. In Section 2 we classify the
observables and discuss our parametrization.  Section 3 contains
phenomenological studies, and finally our conclusions are presented in
Section 4.

\section{Parametrization of Neutrino Observables}

\subsection{Observables and Physical Parameters}

The leptonic superpotential of the minimal supersymmetric model
which implements the seesaw mechanism is
\begin{eqnarray}
\label{w}
W = N^{c}_i (Y_\nu)_{ij} L_j H_2
  -  E^{c}_i (Y_e)_{ij}  L_j H_1 
  + \frac{1}{2}{N^c}_i (M_N)_{ij} N^c_j + \mu H_2 H_1 \,,
\label{suppot}
\end{eqnarray}
where the indices $i,j$ run over three generations and $M_N$
is the heavy singlet-neutrino mass matrix. One can always work in  a basis 
where the charged leptons and
the heavy neutrinos both have real and diagonal mass matrices:
\begin{equation}
(Y_e)_{ij} = {Y}^D_{e_i} \delta_{ij}\,, \;
(M_N)_{ij} = { M_{N_i}} \delta_{ij}\,.
\end{equation}
The matrix $Y_\nu$ contains six physical phases  and  can be 
parametrised as 
\bea
(Y_\nu)_{ij} = Z^\star_{ik} {Y}^D_{\nu_k} X^\dagger_{kj},
\label{Y}
\eea
where $X$ is the analogue of the quark CKM matrix in the lepton sector
and has only one physical phase, and $Z \equiv P_1 \overline{Z} P_2,$
where $\overline{Z}$ is a CKM-type matrix with three real mixing
angles and one physical phase, and
$P_{1,2} \equiv \mrm{diag}(e^{i\theta_{1,3}}, e^{i\theta_{2,4}}, 1 ).$ 
This implies that we have 15 physical parameters in the Yukawa coupling
$Y_\nu$, which together with the 3 unknown heavy masses $M_{N_i}$ make
a total of 18 parameters in the minimal seesaw model~\footnote{The
parameter counting is identical in models with and without 
supersymmetry.}.

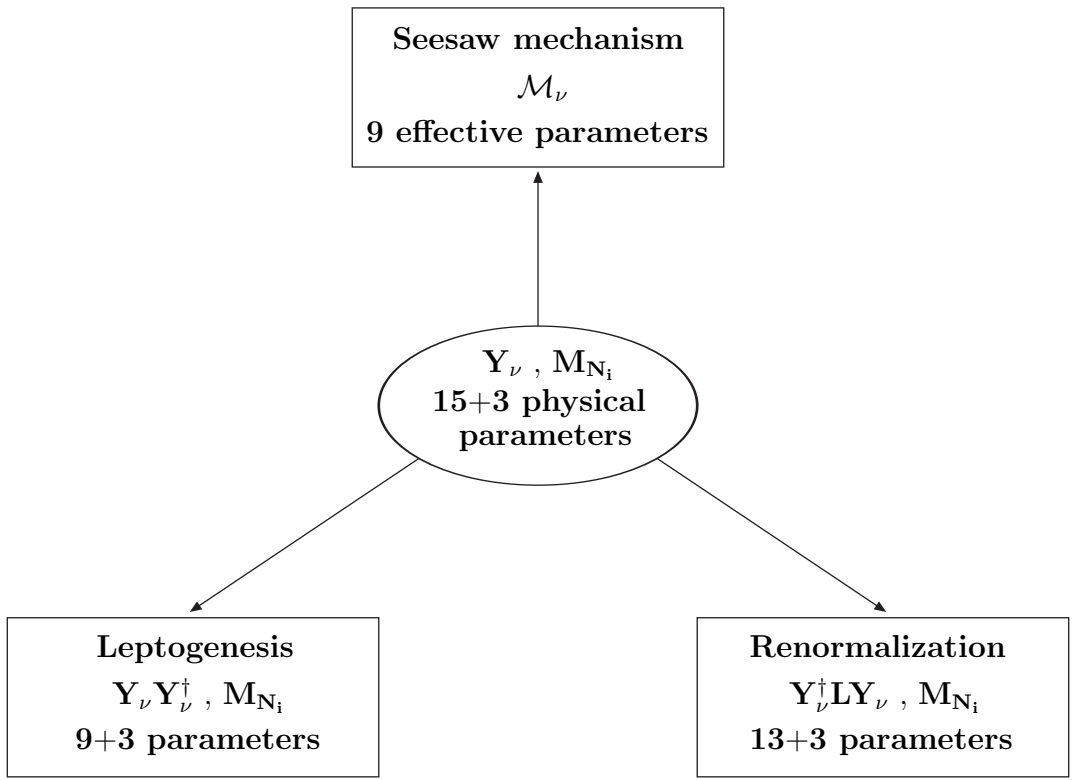
\begin{figure}[t]
\begin{center}
\begin{picture}(400,300)(-200,-150)
\Oval(0,0)(30,60)(0)
\Text(-25,13)[lb]{ ${\bf Y_\nu}$  ,  ${\bf M_{N_i}}$}
\Text(-40,-2)[lb]{{\bf 15$+$3 physical}}
\Text(-30,-15)[lb]{{\bf parameters}}
\EBox(-70,90)(70,150)
\Text(-55,135)[lb]{{\bf Seesaw mechanism}}
\Text(-8,117)[lb]{${\bf {\cal M}_\nu}$}
\Text(-65,100)[lb]{{\bf 9 effective parameters}}
\EBox(-200,-140)(-60,-80)
\Text(-167,-95)[lb]{{\bf Leptogenesis}}
\Text(-165,-113)[lb]{ ${\bf Y_\nu Y_\nu^\dagger}$ , ${\bf M_{N_i}}$}
\Text(-175,-130)[lb]{{\bf 9$+$3 parameters}}
\EBox(60,-140)(200,-80)
\Text(80,-95)[lb]{{\bf Renormalization}}
\Text(95,-113)[lb]{${\bf Y_\nu^\dagger L Y_\nu}$ , ${\bf M_{N_i}}$}
\Text(80,-130)[lb]{{\bf 13$+$3 parameters}}
\LongArrow(0,30)(0,87)
\LongArrow(-45,-20)(-130,-77)
\LongArrow(45,-20)(130,-77)
\end{picture}
\end{center}
\caption{\it
Roadmap for the physical observables derived from $Y_\nu$ and $N_i$.
}
\label{fig0}
\end{figure}

These 18 unknown neutrino parameters give rise to three classes of 
physical observables, as presented diagrammatically in Fig.\ref{fig0}:
\begin{enumerate}

\item[(i)]
Low-energy effective neutrino masses arising from the seesaw mechanism:
\bea 
{\cal M}_\nu={Y}_\nu^T \left({ M_N}\right)^{-1} 
{Y}_\nu v^2 \sin^2\beta .
\label{seesaw1}
\eea
The effective light-neutrino mass matrix ${\cal M}_\nu$ is symmetric and
can be diagonalized by a unitary matrix $U$ as follows:
\bea
U^T {\cal M}_\nu U = {\cal M}^D_\nu\,.
\label{Mnud}
\eea
By a field redefinition, one can rewrite $U \equiv V P_0,$ where
$P_0 \equiv \mrm{diag}(e^{i\phi_1}, e^{i\phi_2}, 1 )$ and $V$ is the MNS 
matrix.
Therefore, all the low-energy neutrino observables such as neutrino 
oscillations, $\beta\beta_{0\nu}$ decay, etc., depend on the 9 effective 
parameters in  ${\cal M}_\nu$, which are functions of all the 18 
parameters in (\ref{suppot}). Whilst neutrino oscillations measure
the mass-squared differences of neutrinos and their mixing angles,
$\beta\beta_{0\nu}$ decay measures one particular combination of
their masses and mixing matrix elements,
\bea
|m_{ee}| \equiv \left|\sum_i U_{ei}^* m_{\nu_i} U_{ie}^\dagger \right| ,
\label{mee} 
\eea
which involves also the Majorana phases.
The NMS mixing phase $\delta$ can in principle be measured
in neutrino oscillations with neutrino factory experiments,
but measurements of the Majorana phases are less 
straightforward \cite{bbth,smirnov,x}.

\item[(ii)]
The idea of baryogenesis via leptogenesis is first to produce a
lepton asymmetry, whose ratio to the entropy density we denote by $Y_L$, 
via the out-of-equilibrium 
decays of the heavy neutrinos $N_i$. This asymmetry 
is converted into the baryon asymmetry of the universe 
via $(B+L)$-violating electroweak sphaleron processes:
\bea
Y_B=\frac{C}{C-1} Y_L \,.
\eea
where $C=8/23$ in the minimal supersymmetric extension of the 
Standard Model (MSSM). The generated lepton asymmetry depends on the initial
neutrino-plus-sneutrino density, the CP asymmetries in neutrino decays,
and on the washout effects. The CP asymmetry produced in 
out-of-equilibrium decays of $N_i$ is given by \cite{vissani}
\begin{eqnarray}
\epsilon_i &=& -\frac{1}{8 \pi} \sum_{l} 
\frac{ \mbox{Im}\Big[
\left( { Y_\nu}{ Y_\nu}^\dagger  \right)^{il}
\left( { Y_\nu}{ Y_\nu}^\dagger \right)^{il}
\Big]}
{ \sum_{j} |{ Y_\nu}^{ij}|^2 }
\nn \\
& &
\sqrt{x_l} \Big[  \mbox{Log} (1+1/x_l) +  \frac{2}{(x_l-1)}\Big] , 
\label{eps}
\end{eqnarray}
where $x_l \equiv (M_{N_l} / M_{N_i})^2.$ It is clear from (\ref{eps})
that the generated asymmetry depends only on the 9 parameters 
(including 3 phases)  in
\begin {equation}
\label{yy+1}
Y_\nu Y_\nu^\dagger 
 = P_1^\star \overline{Z}^\star (Y_\nu^D)^2 \overline{Z}^T P_1\, 
\end{equation}
and on the heavy neutrino masses.

In the case of non-degenerate heavy neutrinos, 
the lepton asymmetry is, to a good approximation, generated only via
the decays of the lightest heavy neutrino $N_1$ (and the corresponding 
sneutrinos), because the very rapid 
washout processes mediated by $N_1$ erase whatever asymmetry had been 
produced previously in $N_{2,3}$ decays \cite{k-sm,p}. 
In this case one has a simple relation
\bea
Y_L= Y_{N_1}^{eq}(0) \,\epsilon_1 \,\kappa_1 \,,
\label{YL}
\eea
where $Y_{N_1}^{eq}(0)$ is the initial thermal equilibrium density
of the lightest neutrinos and $\kappa_1$ is the washout parameter.
The factors multiplying $\epsilon_1,$ in (\ref{YL}) depend on the 
cosmological
scenario. In the case of the standard leptogenesis scenario with thermally 
produced heavy neutrinos one has in thermal equilibrium the initial
condition $Y_{N_1}^{eq}(0)\approx 1/(2 g_*),$ where $g_*\sim 230$ is 
the number of effective degrees of freedom in the MSSM. The washout parameter 
$\kappa_1$ can be precisely calculated by solving the set of
Boltzmann equation for $Y_L$ and for the neutrino density $Y_{N_1},$ 
\bea
&& \frac{\mrm{d}Y_{N_1}}{\mrm{d} z} + \gamma_{N_1}(z) =0\,, \nn \\
&& \frac{\mrm{d}Y_L}{\mrm{d} z} + \gamma_L(z) Y_L + S_L(z) =0\,,
\label{boltz}
\eea
where $Y_L$, $Y_{N_1}$ and the factors $\gamma_{N_1}(z),$ $\gamma_L(z)$ 
and $S_L(z)$ depend on the temperature, which is parametrized by  
$z=M_{1}/T.$
In this scenario, $\gamma_{N_1}(z),$ $S_L(z)$ and $\gamma_L(z)$ are worked out
in great detail \cite{k-sm,bbp} in the Standard Model, for which 
analytical approximations
exist \cite{bcst}, and also in the supersymmetric framework \cite{k-mssm}.
In general, the generated asymmetry depends on four parameters:
the CP asymmetry $\epsilon_1,$ the heavy neutrino mass $M_{N_1},$ and
the lightest light neutrino mass, which determines the overall light-neutrino
mass scale. The effective mass parameter
\bea
\tilde m_1 \equiv \left(Y_\nu Y_\nu^\dagger \right)_{11}
\frac{v^2 \sin^2\beta}{M_{N_1}}
\label{tildem1}
\eea
is also used in discussions of leptogenesis: it depends on the 
10 parameters introduced above. Fixing the baryon asymmetry to agree with 
observation implies a relation
between $\epsilon_1$ and  $\kappa_1$ via (\ref{YL}). Therefore,
only the relations between  $\epsilon_1$ and the masses
$M_{N_1}$ and $m_{\nu_1}$ (or $m_{\nu_3}$ for the inverted mass hierarchy)
are physically relevant.

\item[(iii)]
Renormalization of soft supersymmetry-breaking parameters due to the 
presence of $Y_\nu$ above the heavy-neutrino decoupling scales
induces low-energy processes such as the charged-lepton decays
$\mu\to e\gamma,$ $\mu\to eee,$ $\tau\to l\gamma,$ $\tau\to 3l$, and (in 
the presence of CP violation) EDMs for the electron and muon. In a 
leading-logarithmic approximation,
renormalization modifies the left-slepton mass matrix $m_{\tilde L}$ 
and trilinear soft supersymmetry-breaking $A_e$ terms according to 
\cite{h1}
\begin{eqnarray}
(\delta m_{\tilde{L}}^2)_{ij}&\simeq&
-\frac{1}{8\pi^2}(3m_0^2+A_0^2) (Y^\dagger L Y)_{ij} \,,
\nonumber\\
(\delta A_e)_{ij} &\simeq&
-\frac{1}{8\pi^2} A_0 Y_{e_i} (Y^\dagger L Y)_{ij} \,,
\label{leading}
\end{eqnarray}
which are proportional to
\begin{equation}
Y^\dagger L Y = X Y^D P_2 {\overline Z}^T L {\overline Z}^*
P_2^* Y^D X^\dagger,
\label{YYren}
\end{equation}
where $L$ is a diagonal matrix 
\bea
L_{ij}=\ln (M_{GUT}/M_{N_i}) \delta_{ij}.
\label{L}
\eea
We note that
the CP-violating observables at low energies depend on the leptogenesis
phase in $\overline{Z}.$  
The expression (\ref{YYren}) contains 16 neutrino parameters altogether,
but in completely different combinations from the seesaw mass matrix
(\ref{seesaw1}). Only the two Majorana phases in $P_1$ cancel out in  
(\ref{YYren}). If the heavy neutrinos are exactly degenerate in mass, 
all the renormalization-induced observables are proportional to
$Y_\nu^{\dagger} Y_{\nu} = X (Y_\nu{^D})^2 X^{\dagger}$,
but this is not a good approximation, in general.

The CP-violating observables in LFV processes all depend on a single
CP invariant, which is related to $H =Y_\nu^{\dagger}L Y_{\nu}$ by
$J={\rm Im} H_{12} H_{23} H_{31}$~\cite{lfvcp}. This influences
slepton physics at colliders and also
determines the T-odd asymmetry in $\mu\rightarrow 3e$~\cite{3l}.
The dominant contribution to the lepton EDMs arises 
from threshold corrections to the trilinear coupling $A_e$ due to 
the non-degeneracy of heavy neutrino masses. 
Diagonal phases in $A_e$ are proportional to \cite{ehrs2,ehrs}
\begin{eqnarray}
{\rm Im}[X_j,X_k]_{ii} \log{{M}_{N_k}}/{{M}_{N_j}}&\ne&0,
\label{edm}
\end{eqnarray}
where
$(X_k)_{ij}=
{(Y_\nu^*({M}_{N_k}))_{ki}}
{(Y_\nu({M}_{N_k}))_{kj}}. $
This depends non-trivially
on the CP-violating phases, including the two Majorana phases in
${\cal M}_\nu$ and two phases in $H$ that are irrelevant for LFV.

\end{enumerate}

\subsection{Phenomenological Parametrization of $Y_\nu$ and $M_{N_i}$}

None of the three classes of observables discussed in the previous
subsection, by itself, allows one to measure all the parameters in the
neutrino superpotential (\ref{suppot}).  Moreover, at the moment the only
parameters known experimentally are the 2 effective light neutrino mass
differences and 2 mixing angles, and the baryon asymmetry of the Universe,
whose interpretation requires some cosmological inputs. A central issue
for comprehensive studies of the neutrino sector is how to parametrize
$Y_\nu$ and $M_{N_i}$ in such a way that the effective neutrino parameters
measured in the oscillation experiments are incorporated automatically,
and all other phenomenological constraints are also satisfied.

One may attempt to fix $Y_\nu$ and $M_{N_i}$ in (\ref{suppot}) using model
predictions or some other principle. In that case, however, satisfying the
measured effective neutrino masses and mixings in (\ref{seesaw1}) is a
non-trivial task. One can study the patterns of typical predictions for
LFV processes, EDMs and the baryon asymmetry in any specific model, but
not make comprehensive numerical studies which cover all the allowed
parameter space.

However, as discussed in \cite{di}, in the supersymmetric seesaw model the
low-energy degrees of freedom may in principle be used to reconstruct the 
high-energy neutrino parameters. In~\cite{ehrs} we presented a 
parametrization of 
$Y_\nu$ and $M_{N_i}$ in terms of the light-neutrino mass matrix 
${\cal M}_\nu$ and an auxiliary Hermitian matrix $H$, 
\bea
H=Y_\nu^\dagger D Y_\nu,
\label{H}
\eea
where the diagonal matrix $D$ was chosen in~\cite{ehrs} to be
$D_{ij}=\ln (M_{GUT}/M_{N_i}) \delta_{ij}$, motivated by (\ref{YYren}).
In this case, the parameter matrix $H$ is directly related to the
solutions of the renormalization-group equations (RGEs) for the soft
supersymmetry-breaking slepton masses,
according to (\ref{leading}), and allows us to control the rates for LFV 
processes. Conversely, if any LFV process is observed, one can use 
its value to 
parametrize the heavy neutrino masses and couplings.

In principle the diagonal matrix $D$ can be an arbitrary real matrix.  In
order to study the maximal range for the charged lepton EDMs in the
supersymmetric seesaw model, we also study in this paper the
parametrization with $D=1,$ which is the case discussed in~\cite{di}.
According to (\ref{edm}), the EDMs are not proportional to $H$ in either
case. However, the choice $D=1$ departs from (\ref{edm}) more than the
choice $D=L$. As some of the entries of $H$ must be suppressed in order to
satisfy the stringent experimental upper bound on $\mu\to e\gamma$, one
expects the EDMs to get somewhat larger values if $D=1,$ and this
is indeed the case.

We now present the details of the parametrization.
Starting with (\ref{H}) and using the parametrization~\cite{ci}:
\bea { Y_\nu}= 
\frac{\sqrt{{M_N}} R \sqrt{{\cal M}^D_\nu}\, U^\dagger}{v\sin\beta},
\label{Ynu}
\eea
one can recast (\ref{H}) into a form 
\bea
H'=R^{'\dagger} \overline{M_N} R' ,
\label{meq}
\eea
where $\overline{M_N}$ is a diagonal matrix
\bea
(\overline{M_N})_{ii} = D_{ii} M_{N_i} , 
\label{lneq}
\eea
and
\bea
H'=\sqrt{{\cal M}_\nu^D}^{-1} U^\dagger H U \sqrt{{\cal M}_\nu^D}^{-1}
v^2 \sin^2\beta.
\eea
If (\ref{meq}) can be solved, \ie, if the matrix $H'$ can be 
diagonalized with an orthogonal matrix $R'$, then one can 
solve the heavy neutrino masses from (\ref{lneq}) and calculate the
neutrino Dirac Yukawa couplings from (\ref{Ynu}).
Schematically,
\bea
({\cal M}_\nu \,,\, H) \longrightarrow 
({\cal M}_\nu \,,\overline{M_N},\, R') \longrightarrow 
(Y_\nu\,,\, M_{N_i}) \,,
\label{par}
\eea
where the quantities $\overline{M_N}$ and $R'$ are calculated in
the intermediate step and do not have any independent physical meaning. 
Thus, one has converted the 9 low-energy effective neutrino parameters 
and the 9 free parameters in the Hermitian parameter matrix $H$ into the
18 physical neutrino parameters in $Y_\nu$ and $M_{N_i}.$

The advantages of this parametrization have already been mentioned: it
allows one to control the rates for LFV processes and to scan efficiently
over the allowed parameter space at the same time. The disadvantage of the
parametrization (\ref{par}) is that it is not continuous, because for some
choice of the parameter $H$ there may not exist a matrix $R'$ that
diagonalizes $H'.$ However, in the case of multi-dimensional parameter
spaces such as the one we study, scanning randomly over the allowed
parameters is the most powerful tool, and in practice this disadvantage
does not hinder such a phenomenological study

\section{Phenomenological Analysis}

Using the parametrization developed in the previous section,
we now perform a comprehensive phenomenological
study of all three types of leptonic observables in the 
minimal supersymmetric seesaw model.

We fix the known light neutrino parameters by $\Delta m^2_{32}=3\times
10^{-3}$ eV$^2,$ $\Delta m^2_{21}=4.5\times 10^{-5}$ eV$^2,$
$\tan^2\theta_{23}=1$ and $\tan^2\theta_{12}=0.4$, corresponding to the
LMA solution for the solar neutrino anomaly. Since the experimental
constraint on the angle $\theta_{13}$ is quite stringent, $\sin^2
2\theta_{13}\lsim 0.1$ \cite{CHOOZ,Boehm}, our results depend only weakly
on its actual value. We fix $\sin\theta_{13}=0.1$ and study two cases with
the limiting values of the neutrino mixing phase $\delta=\pi/2$ and
$\delta=0.$ We consider both normally and inversely ordered light neutrino
masses, since neutrino oscillations do not discriminate between them at
present\footnote{Future neutrinoless double beta decay experiments could
resolve this ambiguity~\cite{bbth}.}.

We assume the standard thermal leptogenesis scenario in which the baryon
asymmetry originates only from the lightest heavy neutrino decays, as
described by (\ref{YL}). In our calculations, we solve (\ref{boltz})
numerically using the approximate analytical expressions for the
thermally-averaged interactions given in~\cite{bcst}. 
We start at $T\gg M_{N_1}$ with the initial conditions 
$Y_{N_1}=Y_{N_1}^{eq},$ and  $Y_{L}=0.$ As is appropriate for
supersymmetric models, we concentrate on the low-$M_{N_1}$ regime,
$M_{N_1}\lsim 10^{13}$ GeV, where the approximate solutions for $\kappa_1$
differ from the exact ones~\cite{bbp} by just a constant factor. We
correct our output by this factor to be consistent with~\cite{bbp}. These
results for the washout parameter $\kappa_1$ were derived in the context
of non-supersymmetric seesaw models, but are expected to be a good
approximation also in supersymmetric models, especially for the low $M_{N_1}$
and the moderate
values of $\tan\beta$ that we consider in this work. Because
the heavy neutrinos thermalize very fast, our results are actually
independent of the initial conditions in this low-$M_{N_1}$
regime. Here the wash-out parameter $\kappa_1$ depends practically only on
the effective parameter $\tilde m_1$ (with a small dependence on
$M_{N_1}$). Therefore, with high accuracy, the observed baryon asymmetry
implies via (\ref{YL}) that $\epsilon_1$ and $\tilde m_1$ have a
one-to-one correspondence.

In our subsequent analysis, we require the induced baryon asymmetry 
be in the range~\cite{olive}
\bea
3\times 10^{-11} \lsim Y_B \lsim 9\times 10^{-11},
\label{YB}
\eea
and study its implications on the light and heavy neutrino masses,
the CP asymmetry  $\epsilon_1$, the $\beta\beta_{\nu0}$ decay parameter
$m_{ee},$ the LFV decays and the EDMs of the charged leptons $e$ and 
$\mu$.

As was shown in~\cite{ehrs}, the stringent experimental limit
on $Br(\mu\to e\gamma)$ implies the phenomenological constraints
$H_{12}\ll {\cal O}(1)$ and $H_{13}H_{32}\ll {\cal O}(1)$
on elements of our parameter matrix $H.$ Therefore we 
study two different textures of the matrix $H$:
\bea
H_1=\left(\begin{array}{ccc}
a & 0 & 0 \\
0 & b  & d  \\
0 &  d^\dagger & c
\end{array} \right) \, 
\label{H1}
\eea
and
\bea
H_2=\left(\begin{array}{ccc}
a & 0 & d \\
0 & b  & 0  \\
d^\dagger &  0 & c
\end{array} \right) \, ,
\label{H2}
\eea
where $a,b,c$ are real and $d$ is a complex number.
The texture $H_1$ suppresses $\tau\to e \gamma$ and $d_e$ while 
$\tau\to \mu \gamma$ and $d_\mu$ can be large, and {\it vice versa} for 
$H_2,$
since these processes are sensitive to  $H_{13}$ and $H_{23},$
respectively. 
We consider two forms of the
matrix $H$  \rfn{H}, namely those with $D=L$ and $D=1.$ For the 
textures
$H_1,\,H_2,$ the former suppresses $\mu\to e\gamma$ very efficiently
and all the parameters $a,b,c,|d|$ can be simultaneously of order unity.
However, for the $D=1$ case one of $a,b,c$ must necessarily be
very small in order to keep $\mu\to e\gamma$ below the present 
experimental bound.

In our model, the rates for the LFV processes and EDMs depend on the soft
supersymmetry-breaking parameters. In the following numerical calculations
we fix them at the GUT scale to coincide with one of the post-LEP
benchmark points~\cite{bench} $m_{1/2}=300$ GeV, $m_{0}=100$ GeV,
$A_{0}=-300$ GeV, $\tan\beta=10$ and $sign(\mu)=+1.$ This choice ensures
that all other phenomenological constraints, such as those on the lightest
Higgs boson mass, supersymmetric contributions to $b\to s\gamma$ and
$g_\mu-2,$ cosmological arguments, etc.,  are satisfied within the necessary
accuracy.

As the input parameters we thus have the lightest neutrino mass
$m_{\nu_1}$ (or $m_{\nu_3}$ for inversely-ordered neutrinos), the two
low-scale Majorana phases $\phi_{1,2},$ and the entries $a,b,c,d$ of the
parameter matrices $H_1$ and $H_2.$ We generate these input parameters
randomly in the ranges $(10^{-4}-1)$ eV for $m_{\nu_1}$ (or $m_{\nu_3}$),
$(0-2\pi)$ for $\phi_{1,2},$ and $(10^{-7}-10)$ for $a,b,c,|d|.$ The
distribution for each of them is flat on a logarithmic scale. We require
that $Y_\nu$ remains perturbative up to $M_{GUT}$ and impose the present
constraints on all the LFV processes.

\subsection{Normally-Ordered Light Neutrinos}

First we study the implications on leptogenesis parameters,
$\beta\beta_{\nu0}$ decay, and LFV decays and charged lepton EDMs in the
case of normally-ordered light neutrino masses and mixings. Apart from the
EDMs, the two parametrizations with $D=L$ and $D=1$ give practically
indistinguishable results. Therefore, we present the $D=1$ results only
for the EDMs and the LFV decays, while all the other plots present results
for the parametrization with $D=L$, as in~\cite{ehrs}.

\begin{figure}[htbp]
\centerline{\epsfxsize = 0.5\textwidth \epsffile{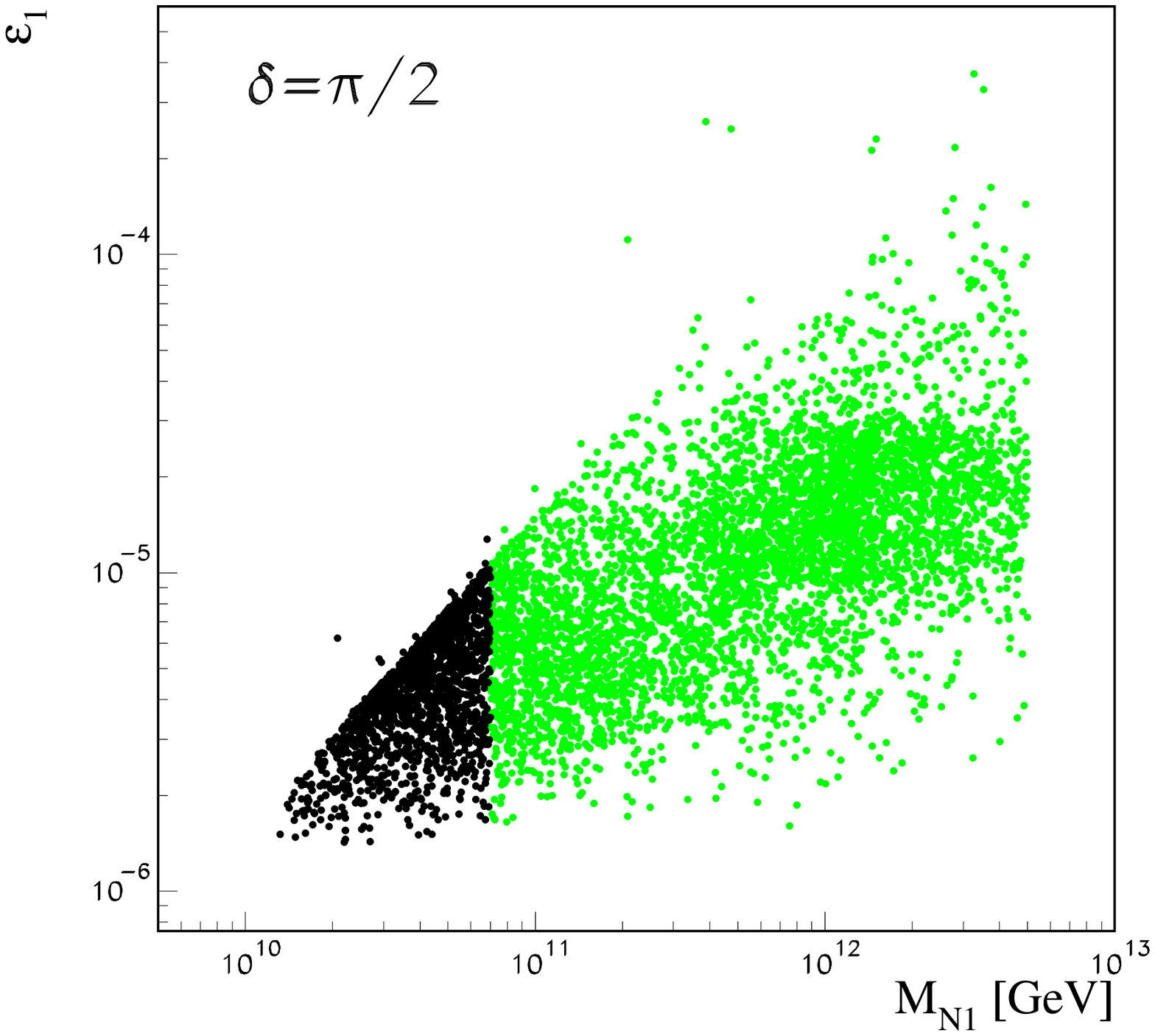} 
\hfill \epsfxsize = 0.5\textwidth \epsffile{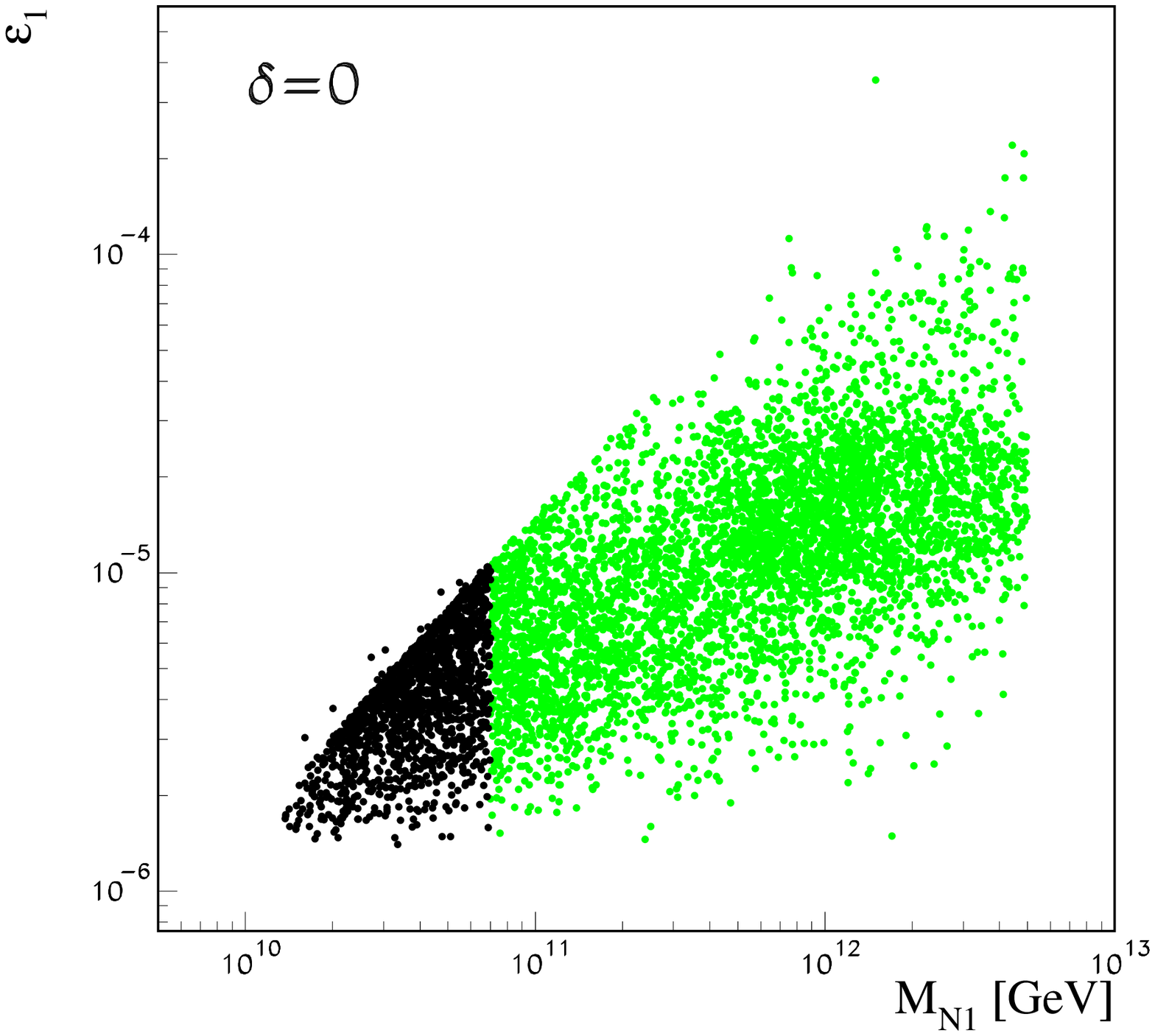} 
}
\caption{\it 
Scatter plot of the CP-violating asymmetry $\epsilon_1$ and the lightest 
heavy neutrino mass $M_{N_1}$ for the two extreme choices of the MNS 
phase $\delta=\pi/2$ and $\delta=0$, assuming
normally-ordered light neutrinos and the texture $H_1.$ 
The baryon asymmetry is required to be in the range (\ref{YB}).
\vspace*{0.5cm}}
\label{fig11h}
\end{figure}

We present in Fig.~\ref{fig11h} scatter plots for the CP-violating
asymmetry $\epsilon_1$ and the lightest heavy neutrino mass $M_{N_1}$, for
two extreme values of the MNS phase: $\delta=\pi/2$ and $\delta=0,$
assuming the texture $H_1.$ Here we also introduce a colour code to study
the distribution of points with different $M_{N_1}$ in the following
plots. The points within a factor of five from the lower bound on
$M_{N_1}$ are black, while the points with larger $M_{N_1}$ are grey
(green).

Fig.~\ref{fig11h} shows immediately that there is no distinction between
the plots for $\delta=\pi/2$ and for $\delta=0.$ In special cases, there
have been studies whether the observed non-zero baryon asymmetry can be
related to the NMS phase $\delta$~\cite{Branco}. Our results imply that
this is not possible in general, and that successful leptogenesis does not
require a non-zero value for $\delta.$

We also see immediately in Fig.~\ref{fig11h} that there is an
$M_{N_1}$-dependent upper bound~\cite{di2} 
on the cosmological CP-violating asymmetry
$\epsilon_1$, and there is a strong lower bound $M_{N_1}\gsim 10^{10}$ GeV
on the $N_1$ mass. This indicates a potential serious conflict with
conventional supersymmetric cosmology which requires an upper bound on the
reheating temperature of the Universe after inflation, derived from
avoiding gravitino overproduction~\cite{gravitino}.

In~\cite{di2}, the analytical bound 
\bea
|\epsilon_1|  \lsim 
\frac{3}{8\pi} \frac{M_{N_1}}{v^2 \sin^2\beta}
(m_{\nu_3}-m_{\nu_1}) ,
\label{epsbound}
\eea
was found in the limit of very hierarchical heavy neutrinos,
$M_{N_1}\ll M_{N_2}\ll M_{N_3}$. 
Our results improve this bound, by including the best available
numerical results for the washout parameter $\kappa_1$ and
allowing for moderately degenerate heavy neutrinos. We have compared
the bound (\ref{epsbound}) with our  numerical calculations. It 
is well satisfied for the low-$M_{N_1}$ points in Fig.~\ref{fig11h},
but can be violated by a large factor for points with high $M_{N_1}$ and
high $\epsilon_1$. In these cases, the heavy neutrinos are
not hierarchical in mass and $\epsilon_1$ is therefore enhanced.
As we see in the next figures, for these points 
the light neutrino masses can also be moderately degenerate.

\begin{figure}[htbp]
\centerline{\epsfxsize = 0.5\textwidth \epsffile{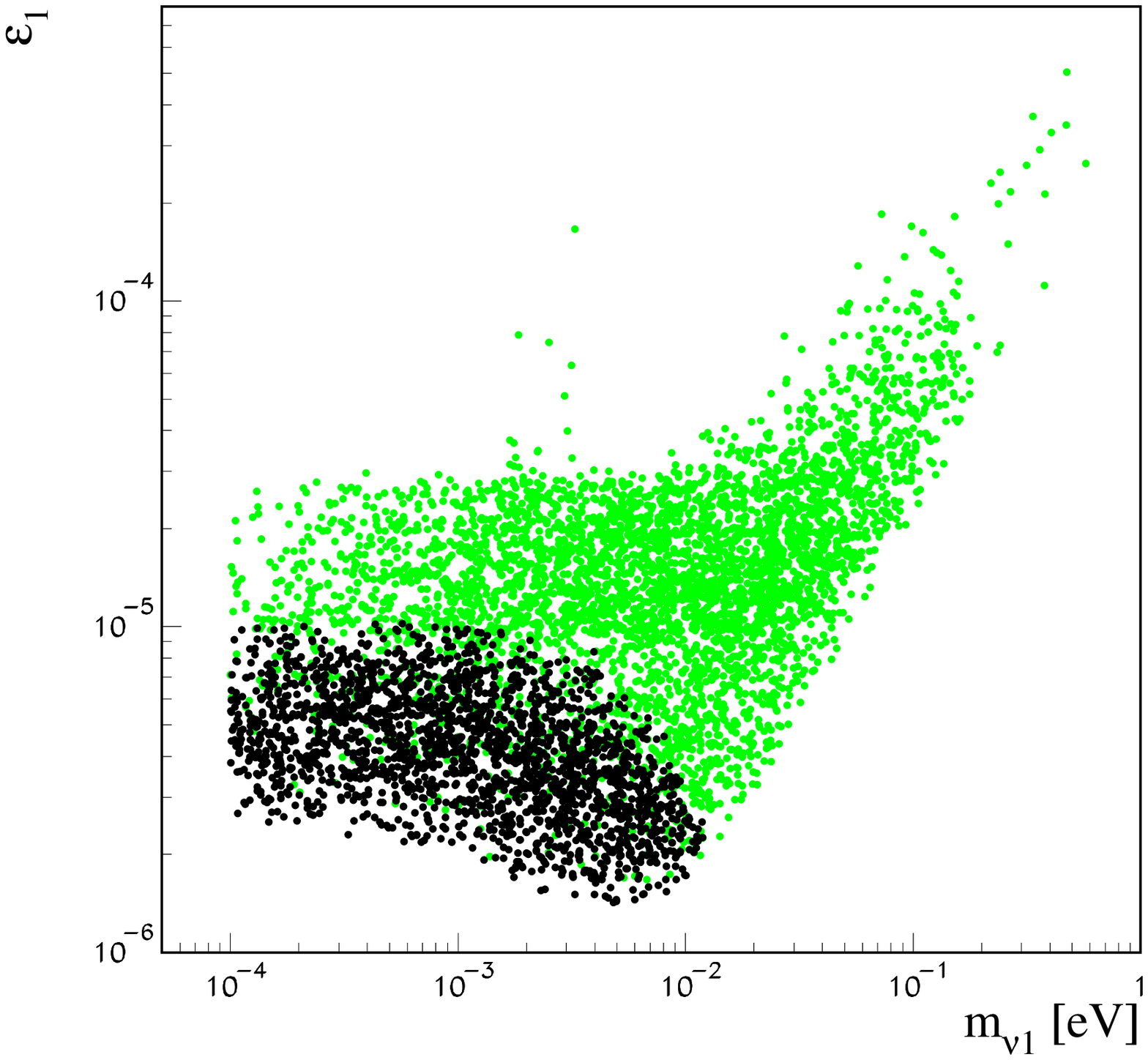} 
\hfill \epsfxsize = 0.5\textwidth \epsffile{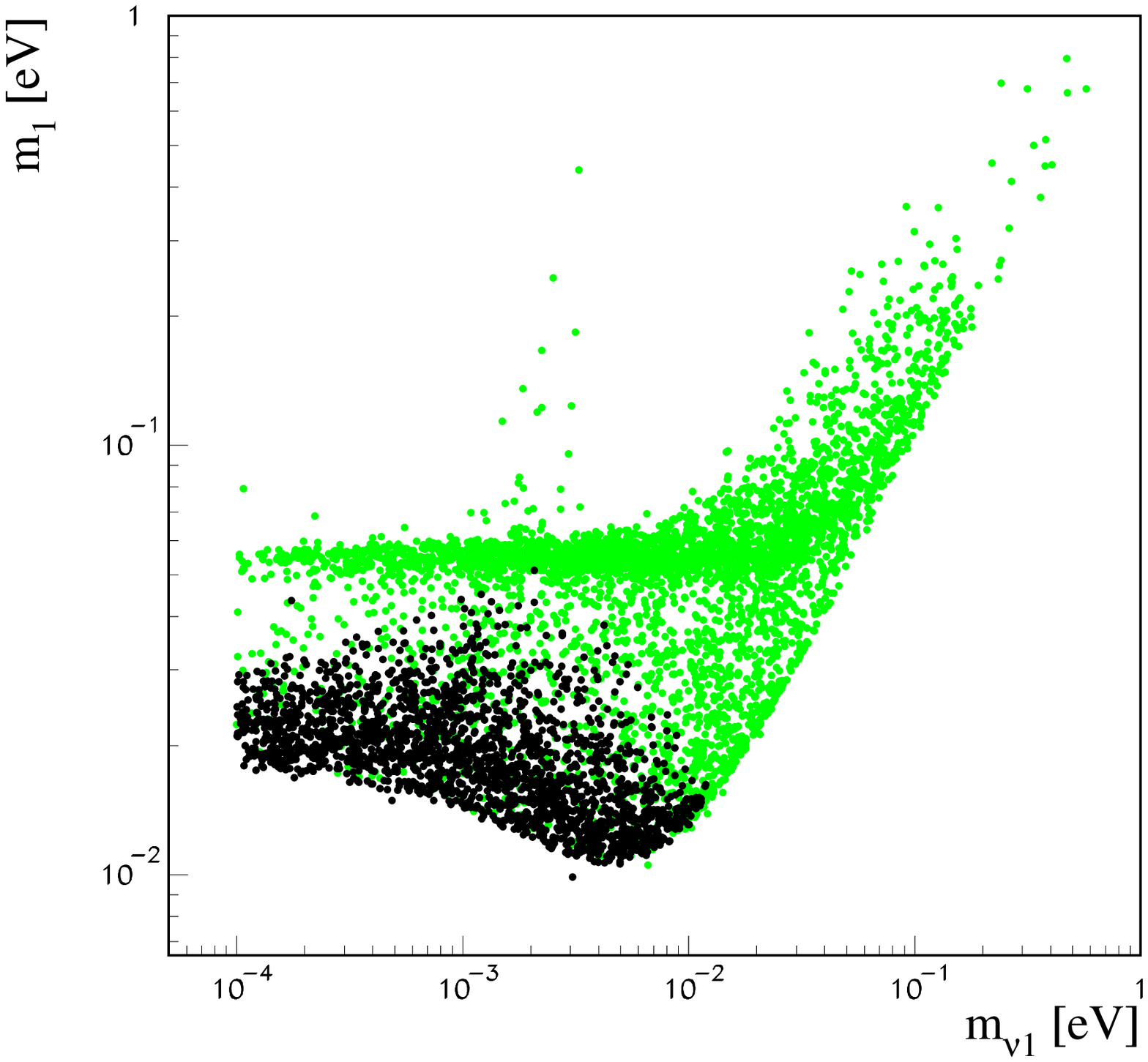} 
}
\caption{\it 
Scatter plot of the CP-violating asymmetry $\epsilon_1$ and the effective 
mass parameter $\tilde m_1$ versus the lightest neutrino mass
$m_{\nu_1}$ for the normal mass hierarchy and the texture $H_1.$
\vspace*{0.5cm}}
\label{fig21h}
\end{figure}

To study correlations between the leptogenesis parameters, we give in
Fig.~\ref{fig21h} scatter plots of the CP-violating asymmetry $\epsilon_1$
and the effective mass parameter $\tilde m_1$ versus the lightest neutrino
mass $m_{\nu_1}$ for the texture $H_1.$ The shape of the both plots is the
same, verifying that, for fixed $Y_B$, the parameters $\epsilon_1$ and
$\tilde m_1$ are not independent parameters in our scenario, as discussed
above. There is no lower bound on $m_{\nu_1}$ in this scenario, and the
upper bound on $m_{\nu_1}$ depends on $M_{N_1},$ as indicated by the
distribution of colours.  The allowed band for $\epsilon_1$ depends only
weakly on $m_{\nu_1}$ for $m_{\nu_1}\lsim 10^{(-1-2)}$ eV, but the
dependence becomes strong for larger $m_{\nu_1}.$ While degenerate light
neutrino masses are disfavoured by our results, confirming the claims
in~\cite{bbp,fhy}, light neutrino masses $m_{\nu_1}\sim {\cal O}(0.1)$ eV
are still perfectly consistent with leptogenesis. Also, notice that for
small $m_{\nu_1}$ the lower bound on $\tilde m_1$ is much stronger than
the limit $\tilde m_1> m_{\nu_1}$ derived in~\cite{bbp,fhy}.

\begin{figure}[htbp]
\centerline{\epsfxsize = 0.5\textwidth \epsffile{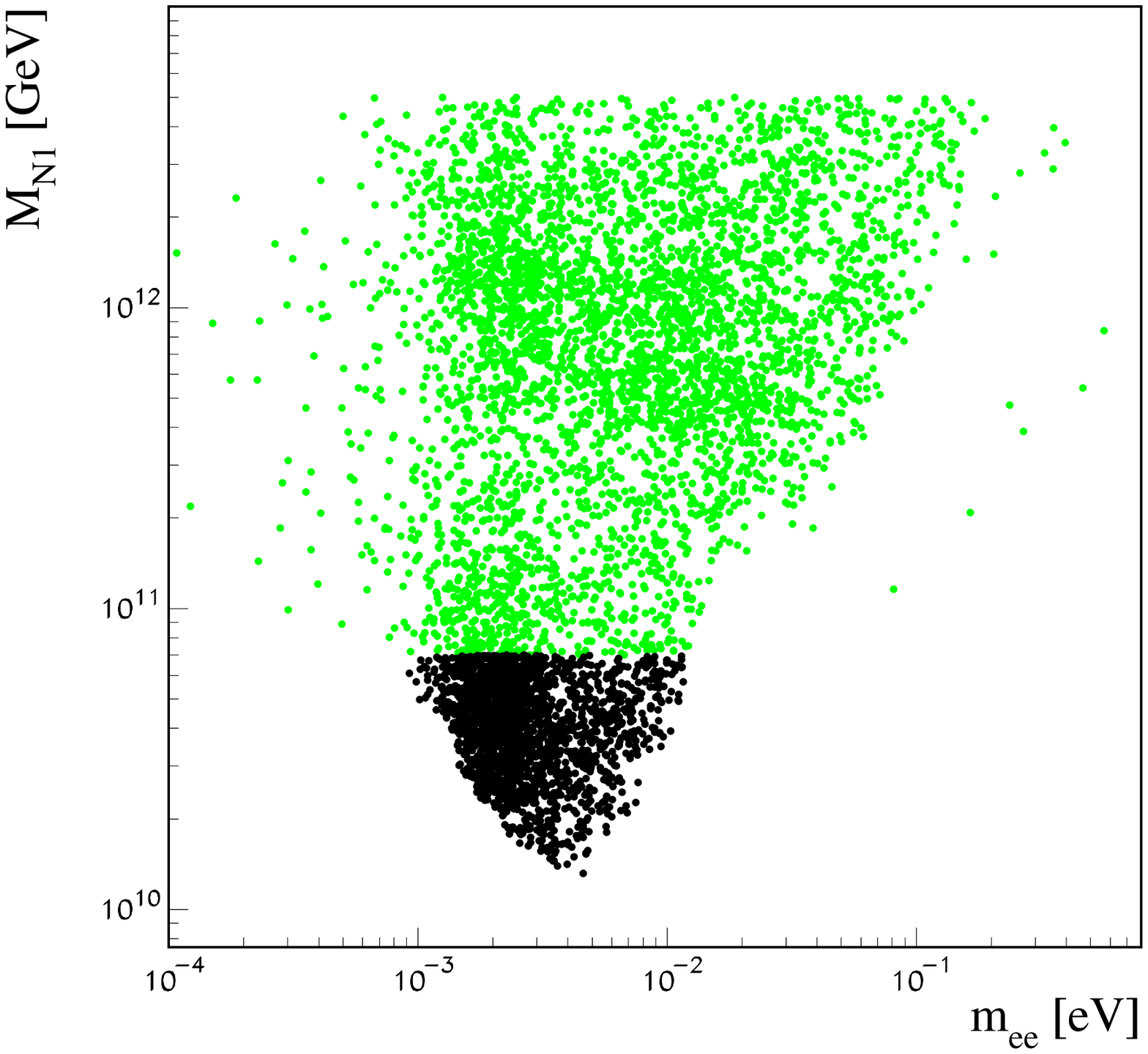} 
\hfill \epsfxsize = 0.5\textwidth \epsffile{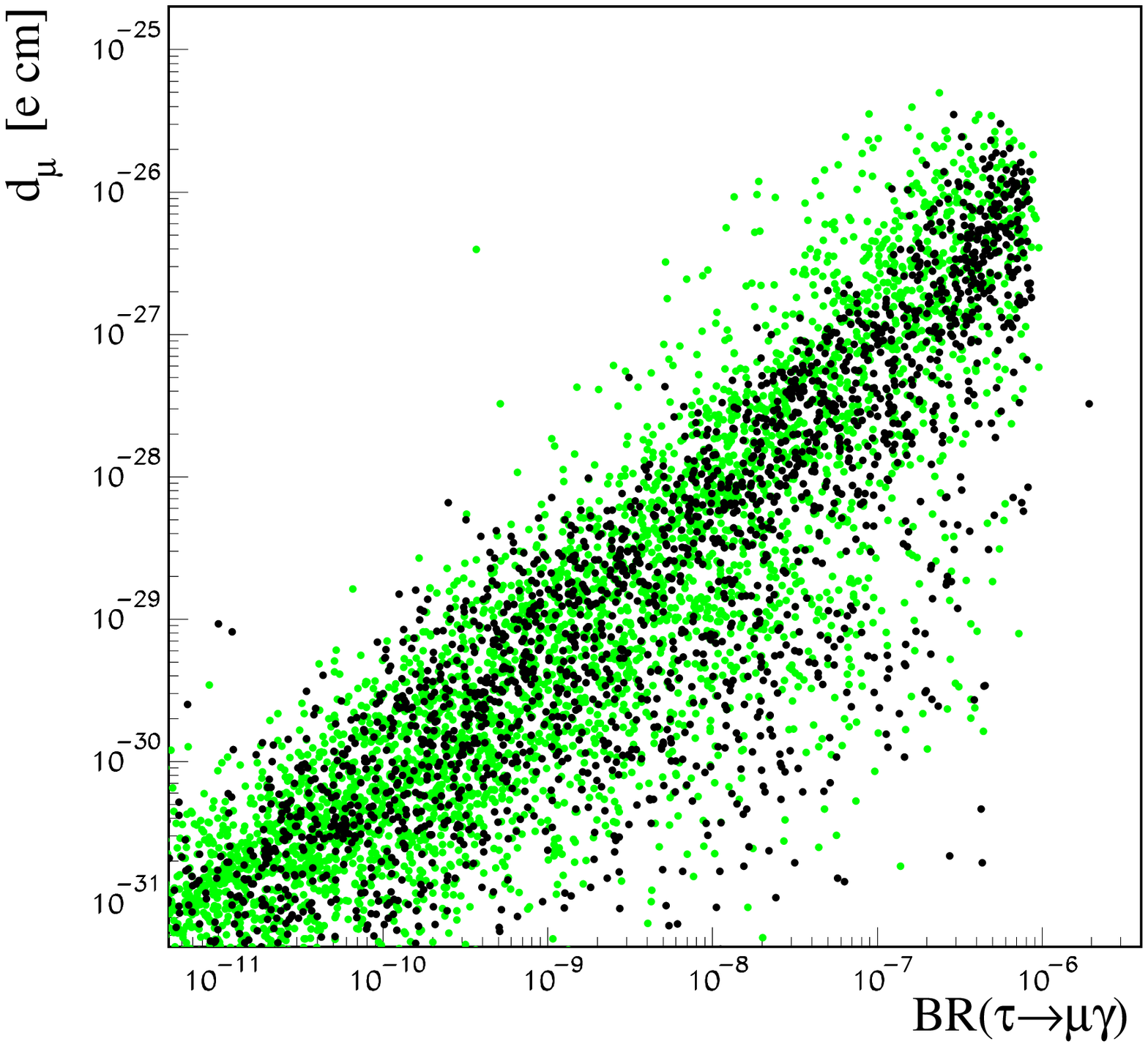} }
\caption{\it 
Scatter plots of
$M_{N_1}$ and the $\beta\beta_{0\nu}$ parameter $m_{ee},$ and of the
muon electric dipole moment $d_\mu$ and the branching ratio for  
$\tau\to\mu\gamma$ decay, for the normal neutrino mass ordering and the
texture $H_1.$
\vspace*{0.5cm}}
\label{fig31h}
\end{figure}

To study the possible implications of the light neutrino masses and
leptogenesis for $\beta\beta_{0\nu }$ decay, the LFV processes and EDMs in
the texture $H_1,$ we give in Fig.~\ref{fig31h} a scatter plot of the
$M_{N_1}$ against the $\beta\beta_{0\nu}$ parameter $m_{ee},$ and the muon
electric dipole moment $d_\mu$ versus the branching ratio for
$\tau\to\mu\gamma.$ Surprisingly, there are both lower and upper limits on
$m_{ee}$, which depend on $M_{N_1}.$ For $M_{N_1}\lsim 10^{11}$ GeV we get
$10^{-3} \lsim m_{ee}\lsim 10^{-2}$ eV.

{\it There is no correlation between leptogenesis and the
renormalization-induced LFV decays and EDMs,} as seen in Fig.~\ref{fig31h}.  
With the chosen soft supersymmetry-breaking parameters,
$Br(\tau\to\mu\gamma)$ can attain the present experimental bound in our
random sample, and the muon EDM may reach $d_\mu\sim 10^{-25}$ e cm. The
former may be observable at the LHC and in B-factory experiments, which
may reach sensitivities $Br(\tau\to\mu\gamma)\sim 10^{(-8-9)}$
\cite{ohshima}, and the latter in experiments at the front end of a
neutrino factory, which may be able to reach $d_\mu\sim 5\times 10^{-26}$
e cm \cite{nufact}. The texture $H_1$ suppresses $Br(\tau\to e\gamma)$ and
$d_e$ below observable limits.

\begin{figure}[htbp]
\centerline{\epsfxsize = 0.5\textwidth \epsffile{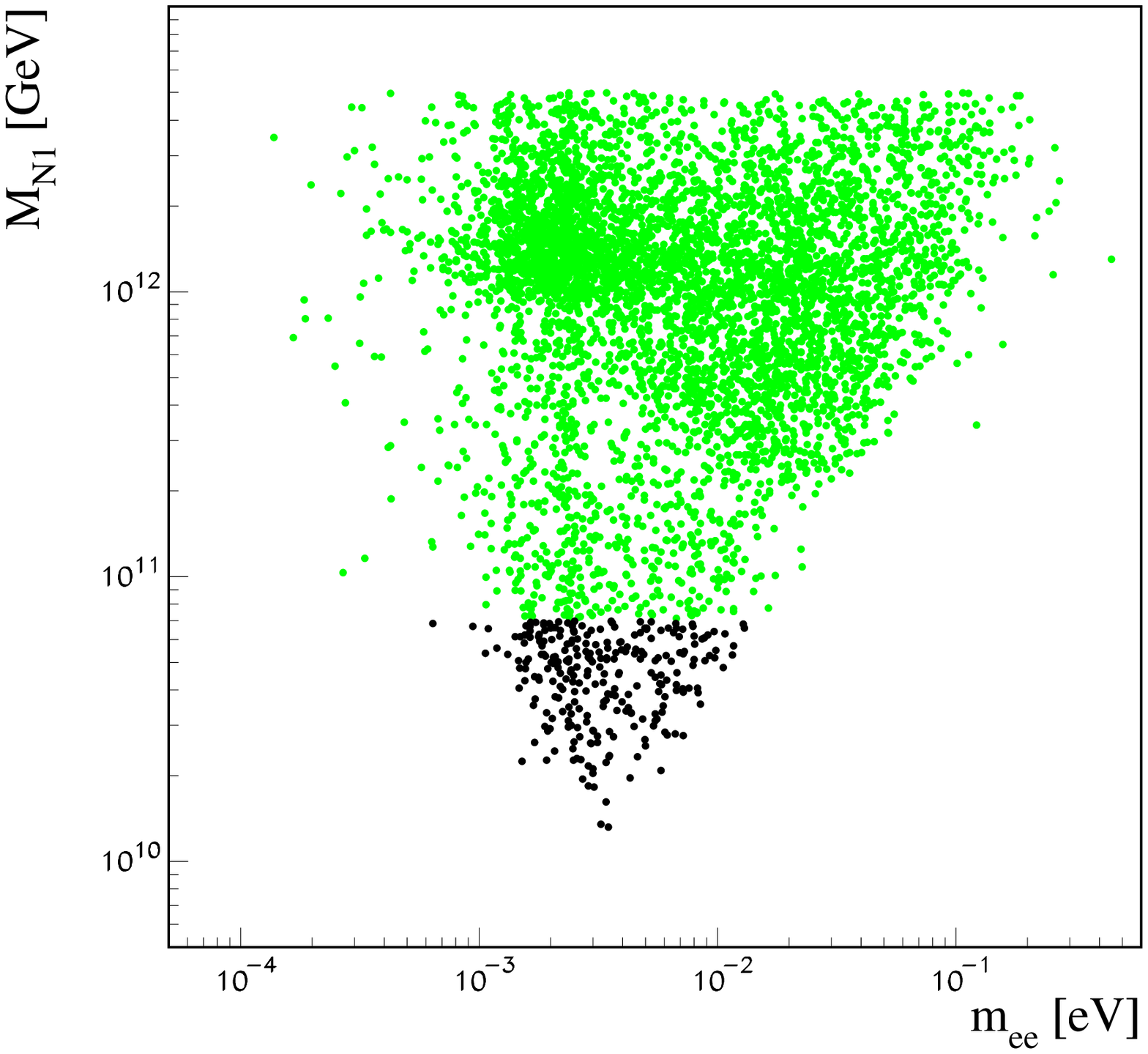} 
\hfill \epsfxsize = 0.5\textwidth \epsffile{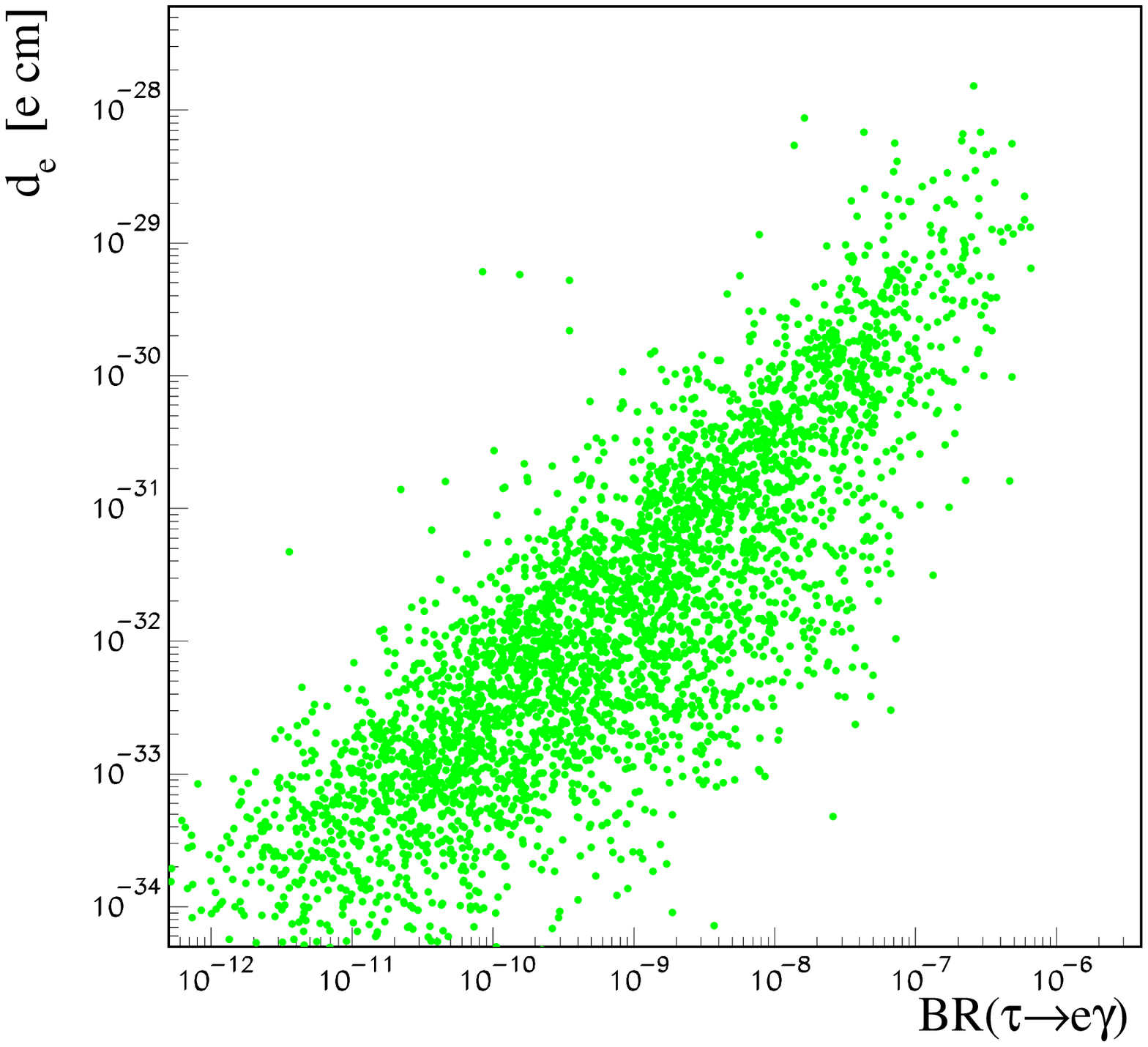} }
\caption{\it 
Scatter plots of
$M_{N_1}$ versus the $\beta\beta_{0\nu}$ parameter $m_{ee},$ and of the
electron electric dipole moment $d_e$ versus the branching ratio for  
$\tau\to e\gamma$, for the normal neutrino mass ordering and the texture 
$H_2.$
\vspace*{0.5cm}}
\label{fig12h}
\end{figure}

We have performed a similar analysis for the texture $H_2.$ The behaviour
of the leptogenesis parameters is almost the same as for the texture
$H_1$, already shown in Figs.~\ref{fig11h} and \ref{fig21h}, so we do not
present further plots here.  The most important differences from $H_1$ can
be seen, however, in Fig.~\ref{fig12h}, where we present scatter plots of
$M_{N_1}$ versus the $\beta\beta_{0\nu}$ parameter $m_{ee},$ and of the
electron electric dipole moment $d_e$ versus the branching ratio for
$\tau\to e\gamma.$ Whilst the lower bound on $M_{N_1}$ is the same as in
the previous case, the distribution of points clearly favours large values
of $M_{N_1}.$ This is a result of a mismatch between the structure of $H_2$
and the light neutrino mass hierarchy $m_{\nu_1}<m_{\nu_2}<m_{\nu_3}.$
Because $H_{11},H_{13}\neq 0$ in $H_2$, the Yukawa couplings for the first
generation tend to be larger than in the case of the texture $H_1.$ As the
lightest ${N_1}$ tends to be related to the lightest light neutrino mass
$m_{\nu_1}$, the seesaw mechanism implies that larger $M_{N_1}$ are
usually needed. Fig.~\ref{fig12h} indicates that this mismatch can be
compensated with a tuning of the input parameters, resulting in a
relatively small number of points at low $M_{N_1}$.

For the texture $H_2$, $Br(\tau\to e\gamma)$ and the electron EDM can be
large, as seen in Fig.~\ref{fig12h}, whilst $Br(\tau\to \mu\gamma)$ and
$d_\mu$ are suppressed below the observable ranges. We find that
$Br(\tau\to e\gamma)$ can be of the same order of magnitude as $Br(\tau\to
\mu\gamma)$, shown in Fig.~\ref{fig31h}.  Importantly, the electron EDM
may exceed $d_e\sim 10^{-28}$ e cm in our numerical examples. This is just
one order of magnitude below the present bound 
$d_e\lsim 1.6\times 10^{-27}$ e cm~\cite{eEDM}. 
As we have not made special attempts to maximize it, it
might even reach larger values with special values of the soft
supersymmetry-breaking parameters.

We see that there are no black points in the right-hand panel of
Fig.~\ref{fig12h}, corresponding to the facts that $Br(\tau\to e\gamma)$
and $d_e$ are very much suppressed. This implies a correlation between the
lower bound on $M_{N_1}$ and the rates of the FLV processes and EDMs. If
$Br(\tau\to e\gamma)$ were to be found at the LHC while $Br(\tau\to
\mu\gamma)$ were not, the lower bound on $M_{N_1}$ from leptogenesis would
be above $10^{11}$ GeV.

\subsection{Inversely-Ordered Light Neutrinos}

\begin{figure}[htbp]
\centerline{\epsfxsize = 0.5\textwidth \epsffile{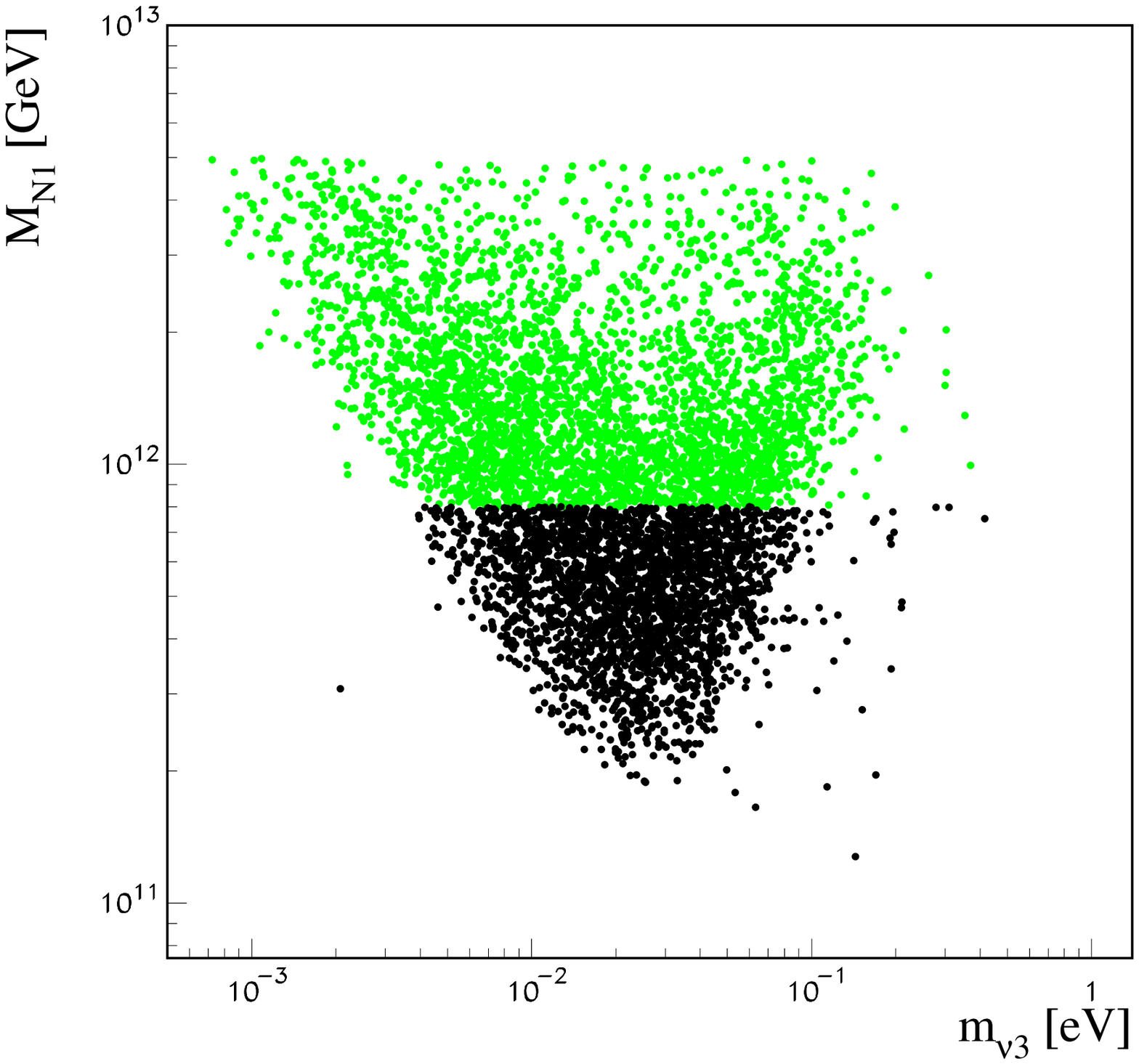} 
\hfill \epsfxsize = 0.5\textwidth \epsffile{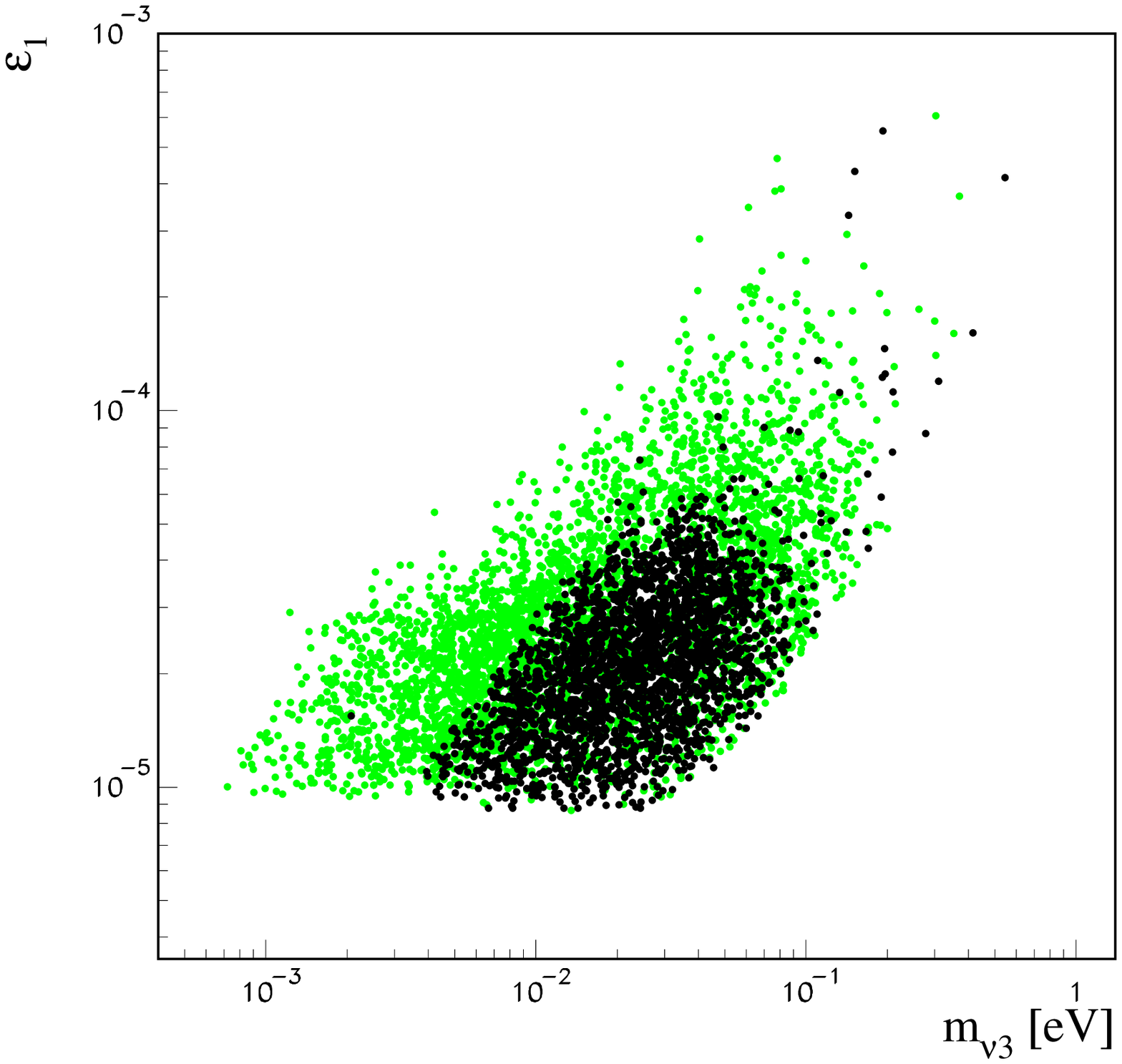} 
}
\caption{\it 
Scatter plots of the lightest heavy neutrino mass $M_{N_1}$ and of the 
CP-violating asymmetry 
$\epsilon_1$ versus $m_{\nu_3}$, for inversely-ordered light 
neutrino masses and the texture $H_2.$
\vspace*{0.5cm}}
\label{fig12i}
\end{figure}
\begin{figure}[htbp]
\centerline{\epsfxsize = 0.5\textwidth \epsffile{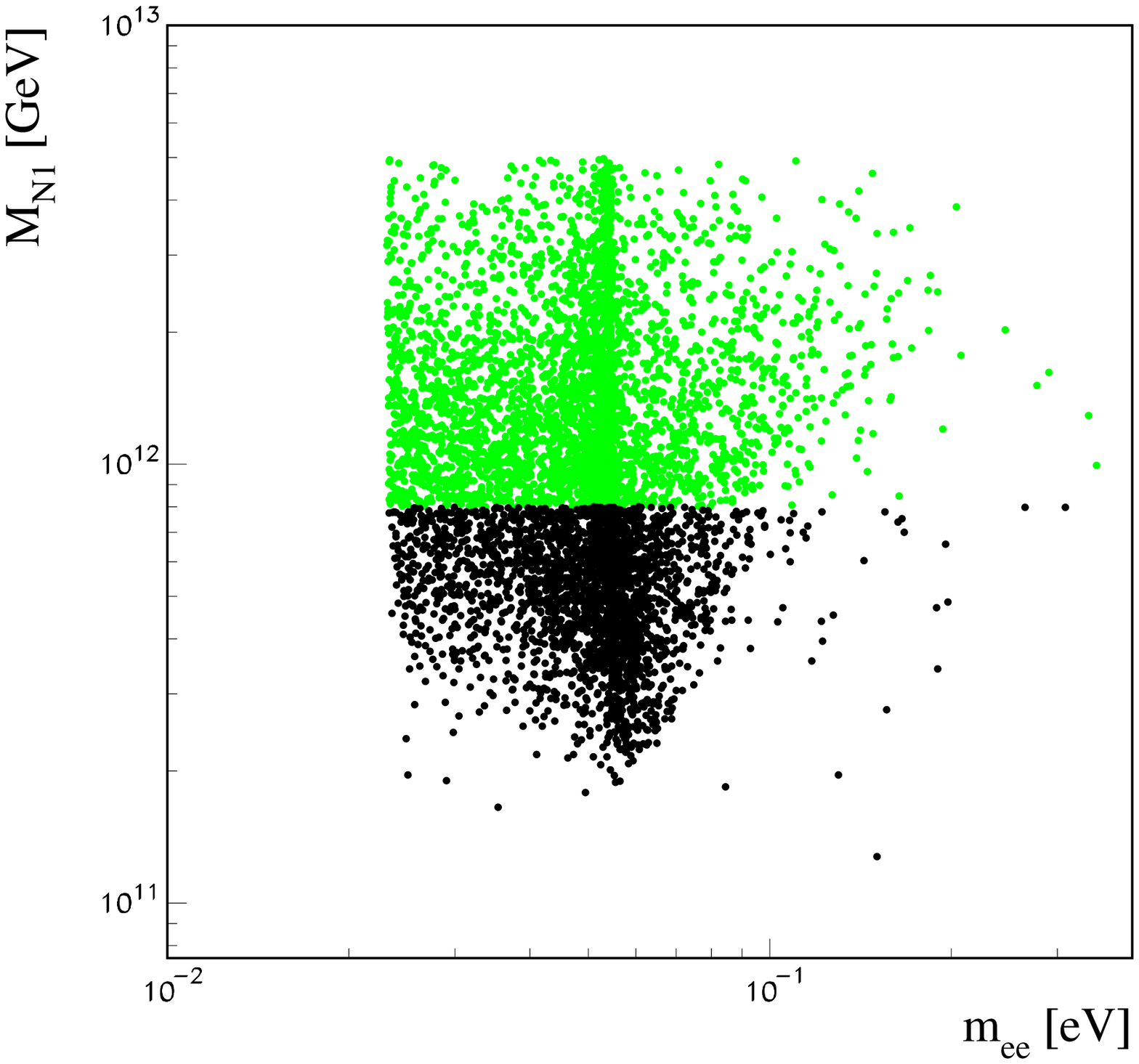} 
\hfill \epsfxsize = 0.5\textwidth \epsffile{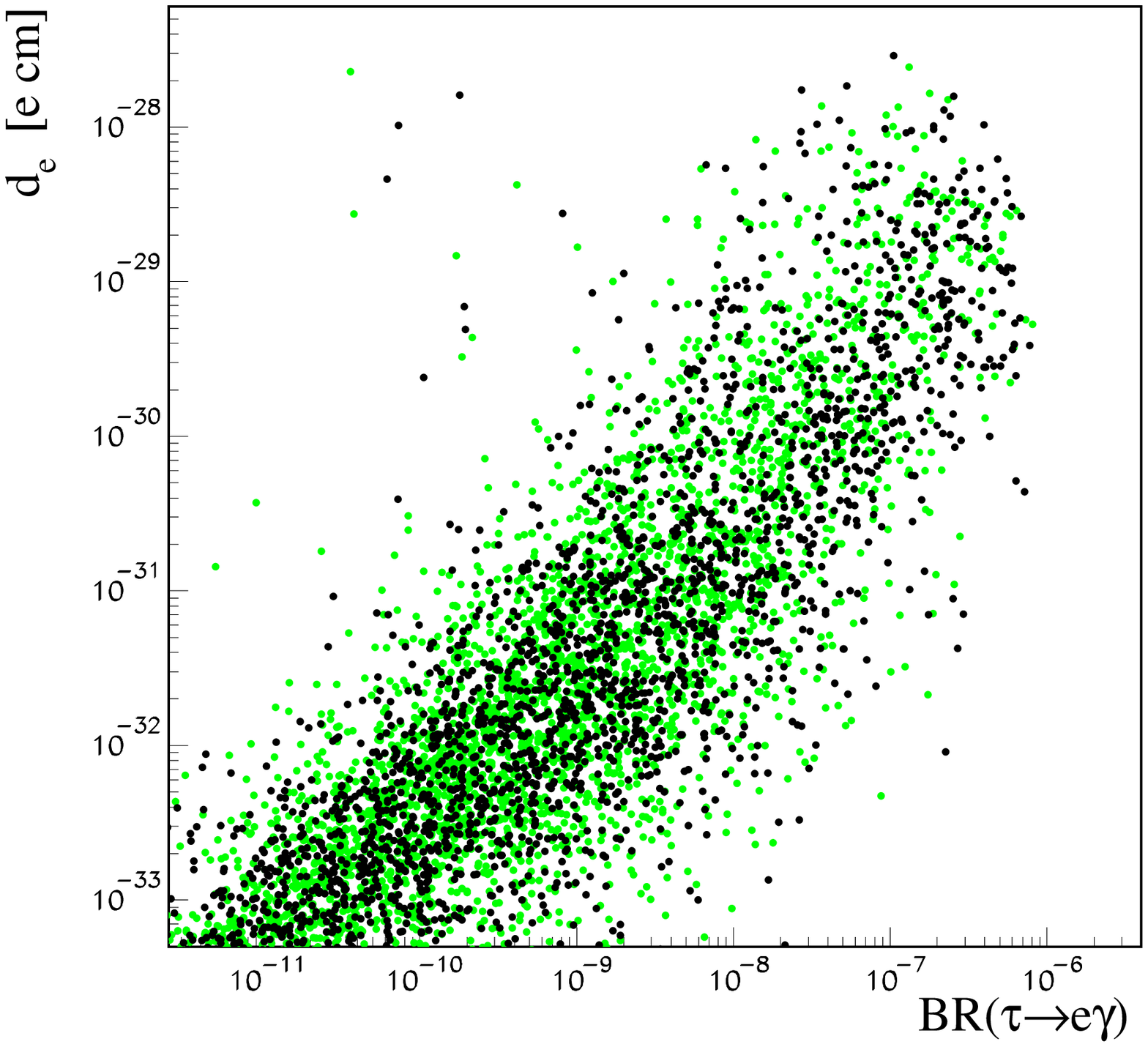} 
}
\caption{\it 
Scatter plots of
$M_{N_1}$ versus $m_{ee}$ and of $d_e$ versus $Br(\tau\to e\gamma)$,
for the inversely-ordered light neutrino masses and the texture $H_2.$
\vspace*{0.5cm}}
\label{fig22i}
\end{figure}

The inverse ordering of light neutrino masses,
$m_{\nu_2}>m_{\nu_1}>m_{\nu_3},$ is still a viable option, since present
neutrino oscillation experiments are not sensitive to the sign of $\Delta
m^2_{32}$. Therefore, it is interesting to study whether this case can be
discriminated from the normal hierarchy of light neutrino masses, using
leptogenesis, LFV and CP-violating processes.

In the inverted light-neutrino mass hierarchy, the correlations between
$\epsilon_1$ and the heavy and light neutrino masses are practically the
same for both textures $H_{1,2}$, and it is sufficient to present results
for just one of them. Since $d_\mu$ is suppressed in $H_{1}$ by two orders
of magnitude compared with the normal hierarchy, due to smaller
third-generation Yukawa couplings, we choose $H_{2}$ for presentation.

We present in Fig.~\ref{fig12i} scatter plots for the lightest heavy
neutrino mass $M_{N_1}$ and the CP-violating asymmetry $\epsilon_1$ versus
the lightest light neutrino mass $m_{\nu_3}.$ As expected, the lower bound
on $M_{N_1}$ is higher than in the normally-ordered case, and is
$M_{N_1}\gsim 10^{11}$ GeV. This result follows from the seesaw mechanism,
since the Yukawa couplings for the first two generations must be larger.
Somewhat surprisingly, leptogenesis implies also a lower bound on the
lightest neutrino mass $m_{\nu_3}$. Therefore the light neutrinos tend to
be degenerate if their masses are inversely ordered. The CP-violating
asymmetry $\epsilon_1$ can be larger than in the previous case, implying
stronger washout effects and somewhat larger light neutrino masses, as
seen in Fig.~\ref{fig12i}.

In Fig.~\ref{fig22i} we give scatter plots of $M_{N_1}$ versus $m_{ee}$
and of $d_e$ versus $Br(\tau\to e\gamma)$ for the same texture as
previously. The sharp lower bound on $m_{ee}$ is the consequence of the
inverted mass ordering. There is a preferred region for $m_{ee}$
determined by $\Delta m^2_{atm}$, and relatively few points extend above
$m_{ee}\gsim {\cal O}(0.1)$ eV. Therefore, even for the inverted mass
hierarchy, observation of $\beta\beta_{0\nu}$ decay in current experiments
is improbable. Again, $d_e$ and $Br(\tau\to e\gamma)$ reach the same
values as in the case of normally-ordered neutrinos, and no strong
correlation between leptogenesis, $d_e$ and $Br(\tau\to e\gamma)$ is
present.

\section{Discussion and Conclusions}

In the context of the minimal supersymmetric seesaw model, we have studied
relations between the light and heavy neutrino masses, thermal
leptogenesis, and LFV decays and EDMs of charged leptons, scanning over
the phenomenologically-allowed parameter space as suggested
in~\cite{ehrs}. There are lower bounds $M_{N_1}\gsim 10^{10}$ GeV and
$M_{N_1}\gsim 10^{11}$ GeV for the normal and inverse hierarchies of light
neutrino masses, respectively. These bounds are in potential conflict with
the gravitino problem in supersymmetric cosmology if the gravitino mass is
below TeV.

In the thermal leptogenesis scenario, one can avoid these bounds by
fine-tuning model parameters, namely the Yukawa couplings $Y_\nu$ and the
heavy neutrino masses $M_N.$ It is well known that the CP-violating
asymmetry may be enhanced if the heavy neutrinos are degenerate in
mass~\cite{p}.  This may allow one to lower~\cite{p,1tev} the lightest
heavy neutrino mass $M_{N_1}$, and so avoid the bounds on the reheating
temperature~\cite{gravitino}. 
Degenerate heavy neutrinos are also consistent with the LMA
solution to the solar neutrino problem~\cite{ricardo}. Because our
parametrization gives $Y_\nu$ and $M_N$ as an output, such fine tunings
cannot be studied by our random scan over the parameter space, and would
require a different approach.

Another way out of the problem would be leptogenesis with non-thermally
produced heavy neutrinos~\cite{nonth,fhy}.  In such a case, the reheating
temperature of the Universe does not limit leptogenesis.  However, they do
not have the same predictivity, and the implications of the observed $Y_B$
on neutrino parameters are lost. It is also possible that gravitino is the
lightest supersymmetric particle, in which case~\cite{bolz} the upper
bound on the reheating temperature is $10^{11}$ GeV and thermal
leptogenesis is possible.

We have found interesting correlations between the heavy and light
neutrino mass parameters. For normally-ordered masses, there is an
$M_{N_1}$-dependent upper bound on $m_{\nu_1}$, whilst for the inverted
hierarchy there are both upper and lower bounds on $m_{\nu_3}.$ Successful
leptogenesis with $m_{\nu_1}\gsim 0.1$ eV is allowed for $M_{N_1}\gsim
10^{12}$ GeV.  There is also an $M_{N_1}$-dependent lower bound on
$m_{ee}$ for normally-ordered light neutrinos, implying its possible
testability in future experiments. On the other hand, $m_{ee}$ has a
preferred value determined by $\Delta m^2_{atm}$ even in the case of the
inverted mass hierarchy. It tends to be below ${\cal O}(10^{-1})$ eV,
making the discovery of $\beta\beta_{0\nu}$ decay in current experiments
improbable.

There is no correlation between leptogenesis, and the LFV decays and 
EDMs of charged leptons. The branching ratios for $\tau\to\mu (e)\gamma$ and
$\mu\to e\gamma$ may saturate their present lower bounds, and the EDMs of
electron and muon reach $d_e\sim 10^{-(27-28)}$ e cm and $d_\mu\sim
10^{-25}$ e cm in our random samples. There is some correlation between
$M_{N_1}$ and the LFV $\tau$ decays for normally-ordered light neutrino
masses: observation of $Br(\tau\to e\gamma)$ would require $M_{N_1}$ to be
an order of magnitude above the lower bound.

The type of parametrization discussed in~\cite{di} and applied
in~\cite{ehrs} and in this work 
is a useful tool for studying the large parameter
space of the minimal supersymmetric seesaw model. In the future, it may
help attempts to devise an experimental strategy for determining 
systematically all the
18 parameters in this sector. It would also be interesting to combine this 
approach with models aiming at predictions for some (all) of the seesaw 
parameters. As we have shown in this paper, this type of parameterization 
can teach us some salutary lessons about leptogenesis and its relations 
to other observables.

\vskip 0.5in
\vbox{
\noindent{ {\bf Acknowledgements} } \\
\noindent  
We thank A. Strumia for discussions, M. Pl\"umacher for a communication,
and J. Hisano and Y. Shimizu for collaboration at early stages.
This work is partially supported by EU TMR
contract No.  HPMF-CT-2000-00460 and ESF grant No. 5135.
}

\end{document}